\documentclass[aps,prb,twocolumn,superscriptaddress,showpacs]{revtex4}
\usepackage{graphicx}

\begin{document}

\title{
Quantum Monte Carlo study on speckle variation due to photorelaxation
of ferroelectric clusters in paraelectric barium titanate
}

\author{Kai Ji}
\email{jikai@post.kek.jp}
\affiliation{
Solid State Theory Division, Institute of Materials Structure Science,
KEK, Graduate University for Advanced Studies, and CREST JST, Oho 1-1,
Tsukuba, Ibaraki 305-0801, Japan
}

\author{Kazumichi Namikawa}
\affiliation{
Department of Physics, Tokyo Gakugei University, Nukuikita 4-1-1, Koganei,
Tokyo 184-8501, Japan
}

\author{Hang Zheng}
\affiliation{
Department of Physics, Shanghai Jiao Tong University, Shanghai 200030, China
}

\author{Keiichiro Nasu}
\affiliation{
Solid State Theory Division, Institute of Materials Structure Science,
KEK, Graduate University for Advanced Studies, and CREST JST, Oho 1-1,
Tsukuba, Ibaraki 305-0801, Japan
}

\date{\today}

\begin{abstract}
Time-dependent speckle pattern of paraelectric barium titanate observed
in a soft x-ray laser pump-probe measurement is theoretically investigated
as a correlated optical response to the pump and probe pulses.
The scattering probability is calculated based on a model with coupled
soft x-ray photon and ferroelectric phonon mode.
It is found that the speckle variation is related with the relaxation
dynamics of ferroelectric clusters created by the pump pulse.
Additionally, critical slowing down of cluster relaxation arises on decreasing
temperature towards the paraelectric-ferroelectric transition temperature.
Relation between critical slowing down, local dipole fluctuation and crystal
structure are revealed by quantum Monte Carlo simulation.
\end{abstract}


\maketitle

\section{Introduction}

Speckle is the random granular pattern produced when a coherent light is
scattered off a rough surface.
It carries information of the specimen surface, for the intensity and
contrast of the speckle image vary with the roughness of surface being
illuminated.\cite{ma77}
Numerous approaches have been devised to identify surface profiles by
either the speckle contrast or the speckle correlation method.\cite{go07}
Recent application of pulsed soft x-ray laser has improved the temporal
and spatial resolution to a scale of picosecond and nanometer.
By this means, dynamics of surface polarization clusters of barium
titanate (BaTiO$_3$) across the Curie temperature ($T_c$) has been
observed,\cite{ta02,ta04} which paves a new way to study the
paraelectric-ferroelectric phase transition.

As a prototype of the ferroelectric perovskite compounds, BaTiO$_3$ undergoes
a transition from paraelectric cubic to ferroelectric tetragonal phase
at $T_c$=395 K.
Below $T_c$, two kinds of ferroelectric domain are developed with mutually
perpendicular polarization.
Structural phase transition and domain induced surface corrugation have
been observed via atomic force microscopy,\cite{ha95} scanning probe
microscopy,\cite{pa98} neutron scattering,\cite{ya69} and polarizing
optical microscopy.\cite{mu96}
In addition to the extensive application of BaTiO$_3$ in technology
due to its high dielectric constant and switchable spontaneous
polarization,\cite{po98} there is also an enduring interest in understanding
the mechanism of paraelectric-ferroelectric phase transition.
It is generally considered that the transition is a classic displacive
soft-mode type driven by the anharmonic lattice dynamics.\cite{ha71,mi76}
However, recent studies have also suggested an order-disorder instability
which coexists with the displacive transition.\cite{za03,vo07}
Therefore, direct observation on creation and evolution of ferroelectric
cluster around $T_c$ is of crucial importance for clarifying the nature
of phase transition.
Since the above-mentioned conventional time-average-based measurements
cannot be adapted to the detection on ultrafast transient status of dipole
clusters, diffraction speckle pattern of BaTiO$_3$ crystal measured by
the picosecond soft x-ray laser has turned out to be an efficient way
for this purpose.

Very recently, Namikawa {\it et al.}\cite{na08} study the polarization
clusters in BaTiO$_3$ at above $T_c$ by the plasma-based x-ray laser speckle
measurement in combination with the technique of pump probe spectroscopy.
In this experiment, two consecutive soft x-ray laser pulses with wavelength
of 160 {\AA} and an adjustable time difference are generated coherently
by the Michelson type beam splitter.
After the photo excitation by the pump pulse, ferroelectric clusters
of nano scale are created in the paraelectric BaTiO$_3$ and tends to be
smeared out gradually on the way back to the equilibrium paraelectric state.
This relaxation of cluster thus can be reflected in the variation of speckle
intensity of the probe pulse as a function of its delay time from the
first pulse.
It has been found that the intensity of speckle pattern decays as
the delay time increases.
Moreover, the decay rate also decreases upon approaching $T_c$, indicating
a critical slowing down of the cluster relaxation time.
Hence, by measuring the correlation between two soft x-ray laser pulses,
the real time relaxation dynamics of polarization clusters in BaTiO$_3$
is clearly represented.
In comparison with other time-resolved spectroscopic study on BaTiO$_3$,
for example the photon correlation spectroscopy with visible laser
beam,\cite{ya08} Namikawa's experiment employs pulsed soft x-ray laser
as the light source.
For this sake, the size of photo-created ferroelectric cluster is reduced
down to a few nanometers, and the cluster relaxation time is at a scale
of picosecond.
This measurement, thus, offers a new insight into the ultrafast quantum
dynamics of domain structure.

In this work, we examine the above-mentioned novel behaviors of ferroelectric
cluster observed by Namikawa from a theoretical point of view, aiming to
provide a basis for understanding the critical nature of BaTiO$_3$.
Theoretically, the dynamics of a system can be adequately described by the
linear response theory, {\it i.e.}, to express the dynamic quantities in
terms of time correlation functions of the corresponding dynamic operators.
In general, the path integral quantum Monte Carlo method is computationally
feasible to handle the quantum many body problems, for it allows the system
to be treated without making any approximation.
However, simulation on real time dynamics with Monte Carlo method is still
an open problem in computational physics because of the formidable numerical
cost of path summation which grows exponentially with the propagation time.
The common approach to circumvent this problem is to perform imaginary
time path integration followed by analytic continuation, and to compute
the real time dynamic quantities using Fourier transformation.
In the present study, the real time correlation functions and real time
dependence of speckle pattern are investigated by this scheme.
Our quantum Monte Carlo simulation demonstrates that the relaxation dynamics
of photo-created nano clusters plays an essential role in determining
the delay time dependence of speckle variation.
Furthermore, it is found that the critical slowing down of photorelaxation
is related to the local dipole fluctuation, which arises near $T_c$ and
stablizes the photo-created ferroelectric cluster.

The remaining of the present paper is organized as follows.
In Sec. II, the model Hamiltonian and theoretical treatment are elaborated.
In Sec. III, our numerical results on speckle correlation and critical
slowing down are discussed in details.
In Sec. IV, a summary with conclusion is presented finally.

\section{Theoretical model and methods}

\subsection{Model Hamiltonian}

The theoretical interpretations for structural phase transition and domain wall
dynamics have be well established in the framework of Krumhansl-Schrieffer
model (also known as $\phi^4$ model).\cite{kr75,au75,sc76,sa02}
In this model, the particles are subject to anharmonic on-site potentials
and harmonic inter-site couplings.
The on-site potential is represented as a polynomial form of the order
parameter such as polarization, displacement, or elasticity, which
displays a substantial change around $T_c$.
Since the $\phi^4$ model is only limited to second-order transitions, in
the present work we invoke a modified Krumhansl-Schrieffer model (also
called $\phi^6$ model)\cite{mo90,kh08} to study the first-order
ferroelectric phase transition of BaTiO$_3$.
In this scenario, the Hamiltonian of BaTiO$_3$ crystal ($\equiv H_f$)
is written as (here we let $\hbar = 1$),
\begin{eqnarray}
H_f &=& -{\omega_0 \over 2} \sum_l {{\partial}^2 \over \partial Q_l^2}
    + U_0 + U_i ,
    \\
U_0 &=& {\omega_0 \over 2} \sum_l \left( Q_l^2 - c_4 Q_l^4 + {c_6 \over 3}
    Q_l^6 \right),
    \\
U_i &=& - {\omega_0 d_2 \over 2} \sum_{\langle l, l' \rangle} Q_l Q_{l'} ,
\end{eqnarray}
where,  $U_0$ and $U_i$ are the on-site potential and inter-site
correlation, respectively.
$Q_l$ is the coordinate operator for the electric dipole moment due to
a shift of titanium ions against oxygen ions, {\it i.e.},
the $T_{1u}$ transverse optical phonon mode.
$\omega_0$ is the dipole oscillatory frequency, $l$ labels the site, and
$\langle l, l' \rangle$ in Eq. (3) enumerates the nearest neighboring pairs.

In order to describe the optical response of BaTiO$_3$ due to x-ray scattering,
we design a theoretical model to incorporate the radiation field and a
weak interplay between radiation and crystal.
The total Hamiltonian reads,
\begin{eqnarray}
H = H_p + H_f + H_{pf} ,
\end{eqnarray}
where
\begin{eqnarray}
H_p = \sum_k \Omega_k a_k^{\dag} a_k, \ \
    \Omega_k = c |k|,
\end{eqnarray}
is the Hamiltonian of polarized light field.
$a_k^{\dag}$ ($a_k$) is the creation (annihilation) operator of a photon
with a wave number $k$ and an energy $\Omega_k$.
$c$ is the light velocity in vacuum.
In Namikawa's experiment, the wave length of x-ray is 160 {\r A}, thus
the photon energy is about 80 eV.
Denoting the odd parity of $T_{1u}$ mode, the photon-phonon scattering
is of a bi-linear Raman type,
\begin{eqnarray}
H_{pf} &=& {V \over N} \sum_{q,q',k} a_{k + {q \over 2}}^{\dag}
    a_{k - {q \over 2}} Q_{q' - {q \over 2}} Q_{-q' - {q \over 2}} ,
\end{eqnarray}
where $V$ is the photon-phonon coupling strength,
$Q_q$ ($\equiv N^{-1/2} \sum_l e^{-i q l} Q_l$) the Fourier component
of $Q_l$ with a wave number $q$.
Without losing generalitivity, here we use a simple cubic lattice, and
the total number of lattice site is $N$.

\subsection{Optical response to pump and probe photons}

\begin{figure}
\includegraphics{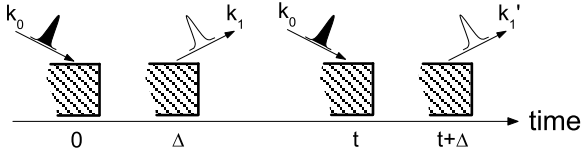}
\caption{
Pulse sequence in an x-ray laser speckle experiment.
The pump and probe pulses of $k_0$ creates and detects ferroelectric clusters
in the sample of paraelectric BaTiO$_3$, respectively, and generate new x-ray
fields in the direction $k_1$ and $k'_1$ after a short time interval $\Delta$.
}
\end{figure}

Since there are two photons involved in the scattering, the photon-phonon
scattering probability can be written as,
\begin{eqnarray}
P(t) = \sum_{k_1, k'_1} \langle \langle
    a_{k_0}(0) a_{k_1}^{\dag}(\Delta) a_{k_0}(t) a_{k'_1}^{\dag}(\Delta + t)
    \nonumber\\
    \times
    a_{k'_1}(\Delta + t) a_{k_0}^{\dag}(t) a_{k_1}(\Delta) a_{k_0}^{\dag}(0)
    \rangle \rangle ,
\end{eqnarray}
where
\begin{eqnarray}
\langle \langle \cdots \rangle \rangle =
\mbox{Tr} (e^{- \beta H} \cdots) / (e^{- \beta H}),
\end{eqnarray}
means the expectation, $\beta$ ($\equiv 1/k_B T$) is the inverse temperature,
and the time dependent operators are defined in the Heisenberg representation,
\begin{eqnarray}
O (t) = e^{i t H} O e^{- i t H} .
\end{eqnarray}
Here, $t$ denotes the time difference between two incident laser pulses as
manifested in Fig. 1, and $k_0$ the wave number of incoming photon.
After a small time interval $\Delta$, the photon is scattered out of
the crystal.
$k_1$ and $k'_1$ are the wave numbers of the first and second outgoing
photons, respectively.

Treating $H_{pf}$ as a perturbation, we separate Hamiltonian of Eq. (4) as,
\begin{eqnarray}
H = H_0 + H_{pf} ,
\end{eqnarray}
where
\begin{eqnarray}
H_0 = H_p + H_f ,
\end{eqnarray}
is treated as the unperturbed Hamiltonian.
By expanding the time evolution operator in Eq. (9) with respect to $H_{pf}$,
\begin{eqnarray}
e^{- i t H} \rightarrow e^{- i t H_0} \left[
    1 - i \int_0^t d t_1 \hat{H}_{pf} (t_1) + \cdots \right] ,
\end{eqnarray}
we find that the lowest order terms which directly depend on $t$ are of
fourth order,
\begin{eqnarray}
P(t) & \rightarrow & \int_0^{\Delta} d t_1 \int_0^{\Delta} d t_2
    \int_0^{\Delta} d t'_1 \int_0^{\Delta} d t'_2
    \sum_{k_1, k'_1}
    \nonumber\\
& & \langle \langle
    a_{k_0} \hat{H}_{pf} (t'_1) e^{i \Delta H_0} a_{k_1}^{\dag}
    e^{i (t - \Delta) H_f} a_{k_0} \hat{H}_{pf} (t'_2)
    \nonumber\\
& & \times e^{i \Delta H_0} a_{k'_1}^{\dag} a_{k'_1} e^{- i \Delta H_0}
    \hat{H}_{pf} (t_2) a_{k_0}^{\dag} e^{- i (t - \Delta) H_f}
    \nonumber\\
& & \times a_{k_1} e^{- i \Delta H_0} \hat{H}_{pf} (t_1) a_{k_0}^{\dag}
    \rangle \rangle ,
\end{eqnarray}
where the operators with carets are defined in the interaction representation,
\begin{eqnarray}
\hat{O} (t) = e^{i t H_0} O e^{- i t H_0} .
\end{eqnarray}

Fig. 2 represents a diagram analysis for this phonon-coupled scattering
process, where photons (phonons) are depicted by the wavy (dashed) lines,
and the upper (lower) horizontal time lines are corresponding to the bra
(ket) vectors.\cite{na94}
Diagram (a) illustrates the changes of wave number and energy of photons
due to the emitted or absorbed phonons.
This is noting but the Stokes and anti-Stokes Raman scattering.
Whereas, diagrams (b)-(e) are corresponding to the exchange, side band,
rapid damping and rapid exchange effects, respectively.

Obviously, diagram (c) brings no time dependence, while diagrams (d) and
(e) only contributes a rapid reduction to the time correlation of two
laser pulses because of the duality in phonon interchange.
In this sense, the time dependence is primarily determined by the diagrams
(a) and (b).
Thus, the scattering probability turns out to be,
\begin{eqnarray}
P (t) &=& \int_0^{\Delta} d t_1 \int_0^{\Delta} d t_2
    \int_0^{\Delta} d t'_1 \int_0^{\Delta} d t'_2
    {2V^4 \over N^4} \sum_{q, q'}
    \nonumber\\
& & \times \langle \langle
    a_{k_0} e^{i t'_1 H_p} a_{k_0}^{\dag} a_{k_0 - q}
    e^{-i (t'_1 - \Delta) H_p} a_{k_0 - q}^{\dag}
    \nonumber\\
& & \times a_{k_0} e^{i t'_2 H_p} a_{k_0}^{\dag} a_{k_0 + q}
    e^{-i (t'_2 - \Delta) H_p} a_{k_0 + q}^{\dag}
    \nonumber\\
& & \times a_{k_0 + q} e^{i (t_2 - \Delta) H_p} a_{k_0 + q}^{\dag}
    a_{k_0} e^{-i t_2 H_p} a_{k_0}^{\dag}
    \nonumber\\
& & \times a_{k_0 - q} e^{i (t_1 - \Delta) H_p} a_{k_0 - q}^{\dag}
    a_{k_0} e^{-i t_1 H_p} a_{k_0}^{\dag}
    \rangle \rangle
    \nonumber\\
& & \times \langle \langle
    \hat{Q}_{q'} (t'_1) \hat{Q}_{q-q'} (t'_1)
    \hat{Q}_{-q+q'} (t + t'_2)
    \nonumber\\
& & \times \hat{Q}_{-q'} (t + t'_2) \hat{Q}_{q'} (t + t_2)
    \hat{Q}_{q-q'} (t + t_2)
    \nonumber\\
& & \times \hat{Q}_{-q+q'} (t_1) \hat{Q}_{-q'} (t_1)
    \rangle \rangle ,
\end{eqnarray}
where the photons and phonons are decoupled, and it becomes evident that
the origin of the $t$-dependence is nothing but the phonon (dipole)
correlation.

Since the photonic part in Eq.(15) is actually time-independent, and
in the case of forward x-ray scattering we have
$|k_0|$$\approx$$|k_1|$$\approx$$|k'_1|$, the normalized probability
can be simplified as,
\begin{eqnarray}
\frac{P(t)}{P(0)} =
    \frac{\sum_{q,q'} | \langle \langle Q_q^2 \rangle \rangle G_{q+q'} (t) |^2}
    {\sum_{q,q'} | \langle \langle Q_q^2 Q_{q+q'}^2 \rangle \rangle |^2} ,
\end{eqnarray}
where
\begin{eqnarray}
G_q(t) &=& -i2 \langle \langle T \hat{Q}_q (t) \hat{Q}_{-q} (0)
    \rangle \rangle ,
\end{eqnarray}
is the real time Green's function of phonon, and $T$ the time
ordering operator.
In deriving Eq. (16), we have also made use of the fact that the light
propagation time in the crystal is rather short.
The Fourier component of Green's function,
\begin{eqnarray}
G_q (\omega) = \int_{- \infty}^{\infty} dt G_q (t) e^{-i \omega t},
\end{eqnarray}
is related to the phonon spectral function [$\equiv A_q (\omega)$]
through,\cite{do74}
\begin{eqnarray}
G_q(\omega) &=& \int_{- \infty}^{\infty} {d \omega' \over 2 \pi}
    {A_q (\omega') \over 1 - e^{- \beta \omega'}} \left(
    {1 \over \omega - \omega' + i0^{+}} \right.
    \nonumber\\
& & \left. -{e^{- \beta \omega'} \over \omega - \omega' - i0^{+}}
    \right) .
\end{eqnarray}
The phonon spectral function describes the response of lattice to the
external perturbation, yielding profound information about dynamic
properties of the crystal under investigation.
Once we get the spectral function, the scattering probability and
correlation function can be readily derived.

\begin{figure*}
\includegraphics{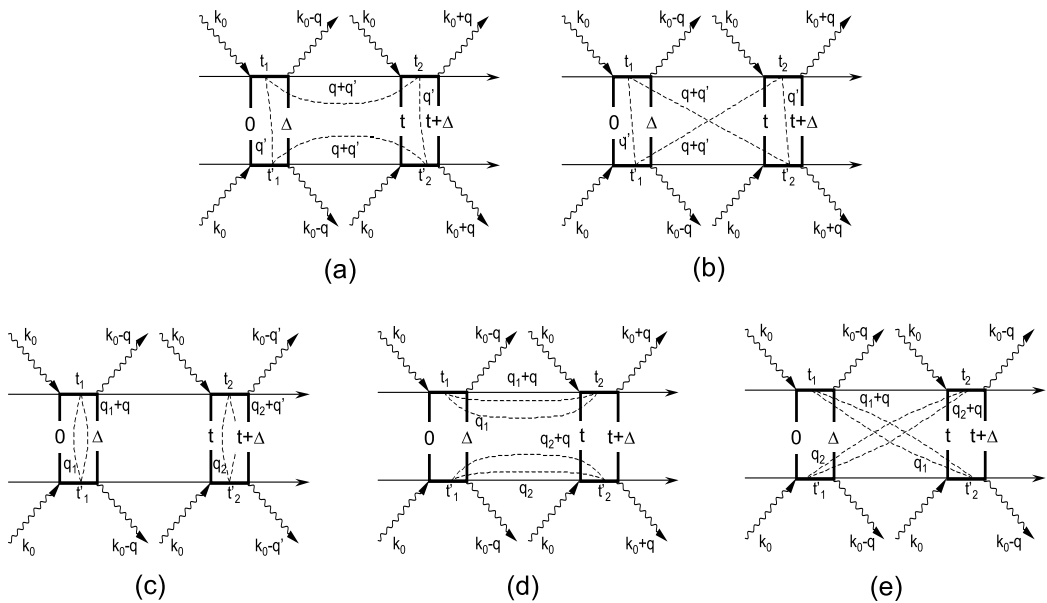}
\caption{
Double-sided Feynman diagrams for scattering process of photon with electric
dipole moment (phonon).
The photons and phonons are denoted by the wavy and dashed lines, respectively.
In each diagram, the upper and lower horizontal time lines represent the
bra and ket vectors, respectively.
}
\end{figure*}

\subsection{Dynamics of crystal}

A mathematically tractable approach to spectral function $A_q (\omega)$
is to introduce an imaginary time phonon Green's function, for it can
be evaluated more easily than its real time counterpart.
In the real space, the imaginary time Green's function is defined as,
\begin{eqnarray}
G_{ll''} (\tau) \equiv -2 \langle \langle T \hat{Q}_l (\tau) \hat{Q}_{l''} (0)
    \rangle \rangle,
\end{eqnarray}
where $\tau$ ($\equiv i t$) is the argument for imaginary time (in this
paper, we follow a convention of using Roman $t$ for real time and Greek
$\tau$ for imaginary time).
The imaginary time dependence of an operator in the interaction
representation is given by
\begin{eqnarray}
\hat{O} (\tau) = e^{\tau H_0} O e^{- \tau H_0} .
\end{eqnarray}
Under the weak coupling approximation, and by using the Suzuki-Trotter
identity, the Green's function can be rewritten into a path integral
form (here we assume $\tau$$>$0),\cite{ji04}
\begin{eqnarray}
G_{ll''} (\tau) = \int {\mathcal D}x e^{-\beta [\Phi_f (x) - \Phi_f]}
    [-2 x_l (\tau) x_{l''} (0)] ,
\end{eqnarray}
where $x_l$ is the eigenvalue of $Q_l$,
\begin{eqnarray}
Q_l |x_l \rangle = x_l |x_l \rangle,
\end{eqnarray}
$\Phi_f (x)$ is the path-dependent phonon free energy
\begin{eqnarray}
e^{- \beta \Phi_f (x)} = e^{- \int_0^\beta d \tau \Omega_f [x(\tau)]} ,
\end{eqnarray}
with
\begin{eqnarray}
\Omega_f &=& \sum_l \left[
    {1 \over 2 \omega_0} \left( {\partial x_l \over \partial \tau} \right)^2
    + {1 \over 2} \omega_0 x_l^2 - {1 \over 2} \omega_0 c_4 x_l^4
    \right.
    \nonumber \\
& & \left. + {1 \over 6} \omega_0 c_6 x_l^6 \right]
    -{1 \over 2} \omega_0 d_2 \sum_{\langle l,l' \rangle} x_l x_{l'} ,
\end{eqnarray}
and $\Phi_f$ is the total phonon free energy,
\begin{eqnarray}
e^{- \beta \Phi_f} = \int {\mathcal D}x e^{- \beta \Phi_f (x)}.
\end{eqnarray}
In the path integral notation, the internal energy of crystal $E_f$
($\equiv \langle \langle H_f \rangle \rangle$) is represented as
\begin{eqnarray}
E_f &=& \int {\mathcal D}x e^{-\beta [\Phi_f (x) - \Phi_f]}
    \left[ \omega_0 \sum_l \left(x_l^2
    - {3 \over 2} c_4 x_l^4 \right. \right.
    \nonumber \\
& & \left. \left. +{2 \over 3} c_6 x_l^6 \right)
    - \omega_0 d_2 \sum_{\langle l,l' \rangle} x_l x_{l'} \right] ,
\end{eqnarray}
from which the heat capacity can be derived as
\begin{eqnarray}
C_f^V = \left( {\partial E_f \over \partial T} \right)_V .
\end{eqnarray}

The Green's function of momentum space is given by,
\begin{eqnarray}
G_q (\tau) = {1 \over N} \sum_{l,l''} e^{i q (l - l'')} G_{ll''} (\tau),
\end{eqnarray}
which is connected with the phonon spectral function $A_q (\omega)$
through\cite{bo96}
\begin{eqnarray}
G_q (\tau) = - \int_0^{\infty} {d \omega \over 2 \pi}
    \frac{\cosh \left[ \left( {1 \over 2} \beta - \tau \right) \omega \right]}
    {\sinh \left( {1 \over 2} \beta \omega \right)}
    A_q (\omega) .
\end{eqnarray}
Solving this integral equation is a notoriously ill-posed numerical problem
because of the highly singular nature of the kernel.
In order to analytically continue the imaginary time data into real frequency
information, specialized methods are developed, such as maximum entropy
method\cite{sk84} and least squares fitting method.\cite{ya03}
In present work, we adopt the iterative fitting approach,\cite{ji04}
for it can give a rapid and stable convergence of the spectrum without
using any prior knowledge or artificial parameter.
Since the phonon spectral function does not yield a specific sum rule
like the case of electron, here we introduce an auxiliary spectral function
$\tilde A_q (\omega)$ which is defined by
\begin{eqnarray}
\tilde{A}_q (\omega) \equiv
    - \frac{\coth \left( {1 \over 2} \beta \omega \right)}{G_q (\beta)}
    A_q (\omega).
\end{eqnarray}
Substituting $\tilde A_q (\omega)$ into Eq. (30), we get
\begin{eqnarray}
G_q (\tau) = \int_0^{\infty} {d \omega \over 2 \pi}
    \frac{\cosh \left[ \left( {1 \over 2} \beta - \tau \right) \omega \right]}
    {\cosh \left( {1 \over 2} \beta \omega \right)} G_q (\beta)
    \tilde{A}_q (\omega) .
\end{eqnarray}
It can be easily shown that this auxiliary spectral function satisfies
a sum rule,
\begin{eqnarray}
\int_0^{\infty} {d \omega \over 2 \pi} \tilde{A}_q (\omega) = 1 ,
\end{eqnarray}
which allow us to solve the integral equation of Eq. (32) by
the iterative fitting approach.
Once $\tilde{A}_q (\omega)$ is reproduced, the phonon spectral function
$A_q (\omega)$ can be obtained from Eq. (31).

\section{Numerical results and discussion}

\subsection{Optical responses}

Based on the path integral formalisms, the imaginary time Green's function
can be readily calculated via a standard quantum Monte Carlo
simulation.\cite{ji04}
Our numerical calculation is performed on a 10$\times$10$\times$10 cubic
lattice with a periodic boundary condition.
The imaginary time is discretized into 10-20 infinitesimal slices.
As already noticed for the analytic continuation,\cite{gu91} if the imaginary
time Green's function is noisy, the uncertainty involved in the inverse
transform might be very large, and the spectral function cannot be determined
uniquely.
In order to obtain accurate data from quantum Monte Carlo simulation,
a hybrid algorithm\cite{ji04} has been implemented in our calculation.
Besides, we pick out each Monte Carlo sample after 100-200 steps to reduce
the correlation between adjacent configurations.
The Monte Carlo data are divided into 5-10 sets, from which the 95\% confidence
interval is estimated through 10,000 resampled set averages by the percentile
bootstrap method.
We found that about 1,000,000 Monte Carlo configurations are sufficient to
get well converged spectral functions and real time dynamic quantities.

\begin{figure}
\begin{center}
\includegraphics{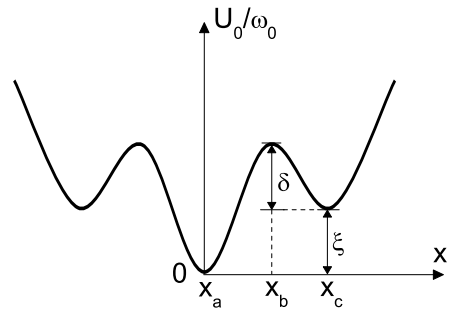}
\end{center}
\caption{
On-site potential $U_0$ for the modified Krumhansl-Schrieffer model in
the unit of $\omega_0$.
The coordinates of potential extrema are denoted by $x_a$, $x_b$, and $x_c$.
$\delta$ and $\xi$ are two parameters adopted to characterize this potential.
}
\end{figure}

In the numerical calculation, the phonon frequency $\omega_0$ is assumed
to be 20 meV,\cite{zh94} the inter-site coupling constant $d_2$ is fixed
at a value of 0.032, whereas $c_4$ and $c_6$ are selected to make the
on-site $U_0$ a symmetric triple-well potential.
As shown in Fig. 3, this triple-well structure is featured by five potential
extrema located at $x_a$, $\pm x_b$ and $\pm x_c$, where
\begin{eqnarray}
x_a &=& 0 , \\
x_b &=& \sqrt{ \frac{c_4 - \sqrt{ c^2_4 - c_6 }}{c_6} } , \\
x_c &=& \sqrt{ \frac{c_4 + \sqrt{ c^2_4 - c_6 }}{c_6} } .
\end{eqnarray}

\begin{figure}
\begin{center}
\includegraphics{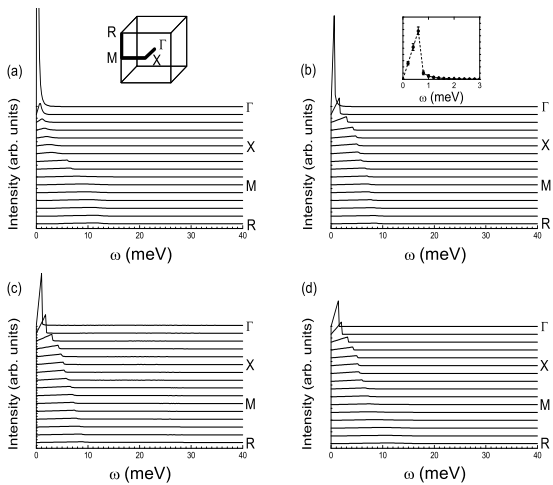}
\end{center}
\caption{
Phonon spectral function along the line $\Gamma$XMR of Brillouin zone
in the paraelectric phase at various temperatures: (a) 1.001$T_c$,
(b) 1.012$T_c$, (c) 1.059$T_c$, and (d) 1.176$T_c$, where $T_c$=386 K.
The inset of panel (a) shows the Brillouin zone with high symmetry lines.
The inset of panel (b) represents the spectrum at $\Gamma$ point
when $T$=1.012$T_c$.
Error bars mark the 95\% confidence interval.
}
\end{figure}

In Figs. 4 and 5, we show the optical responses of crystal, where
$c_4$=2.0132$\times$10$^{-2}$ and $c_6$=3.2595$\times$10$^{-4}$ are used.
Fig. 4 presents the phonon spectral functions in the paraelectric phase
at different temperatures: (a) $T$=1.001$T_c$, (b) $T$=1.012$T_c$, (c)
$T$=1.059$T_c$ and (d) $T$=1.176$T_c$, where $T_c$=386 K.
In each panel, the spectra are arranged with wave vectors along the $\Gamma$XMR
direction of Brillouin zone [see in the inset of panel (a)], and $\omega$
refers to energy.
In the inset of panel (b), the spectrum at $\Gamma$ point for $T$=1.012$T_c$
is plotted with 95\% confidence interval illustrated by the error bars.
Since the spectra are symmetric with respect to the origin $\omega$=0,
here we only show the positive part of them.
In Fig. 4, when the temperature decreases towards $T_c$, as already well-known
for the displacive type phase transition, the energy of phonon peak is
gradually softened.
In addition, a so-called central peak, corresponding to the low energy
excitation of ferroelectric cluster, appears at the $\Gamma$ point.
The collective excitation represented by this sharp resonant peak is
nothing but the photo-created ferroelectric cluster.
On decreasing temperature, spontaneous polarization is developed
locally as a dipole fluctuation in the paraelectric phase.
This fluctuation can stabilize the photo-created ferroelectric cluster,
leading to a dramatically enhanced peak intensity near $T_c$.

The appearance of sharp peak at $\Gamma$ point nearby $T_c$ signifies a long
life-time of the photo-created ferroelectric clusters after irradiation.
Thus, near $T_c$, they are more likely to be probed by subsequent laser
pulse, resulting in a high intensity of speckle pattern.
Keeping this in mind, we move on to the results of scattering probability.
In Fig. 5, we show the variation of normalized probability $P(t)/P(0)$ as a
function of $t$ (time interval between the pump and probe photons).
Temperatures for these curves correspond to those in the panels (a)-(d)
of Fig. 4, respectively.
In this figure, $P(t)/P(0)$ declines exponentially, showing that the speckle
correlation decreases with $t$ increases as a result of the photorelaxation
of ferroelectric cluster.
When $t$ is long enough, the crystal returns to the equilibrium
paraelectric state.
In addition, as shown in the figure, the relaxation rate bears
a temperature dependence.
On approaching $T_c$, the duration for return is prolonged, indicative
of a critical slowing down of the relaxation.
This is because with the decrease of temperature, the fluctuation of local
polarization is enhanced, and a long range correlation between dipole moments
is to be established as well, making the relaxation of photo-created
clusters slower and slower.

\begin{figure}
\begin{center}
\includegraphics{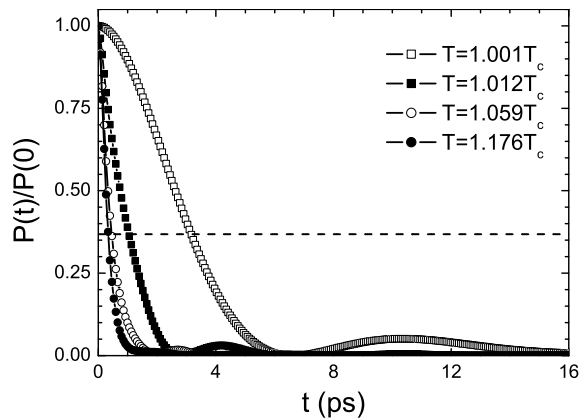}
\end{center}
\caption{
Normalized speckle scattering probability as a function of time for
paraelectric BaTiO$_3$, at various temperatures.
Horizontal dashed line denotes $P(t) = P(0)/e$.
}
\end{figure}

\subsection{Critical slowing down of photorelaxation}

In order to quantitatively depict the critical slowing down, we introduce
a relaxation time $t_r$ to estimate the time scale of relaxation, which
is the time for $P (t)$ to be reduced by a factor of $e$ from $P (0)$.
In Fig. 5, $P (t)$=$P (0)/e$ is plotted by a horizontal dashed line.
Correspondingly, $t_r$ is the abscissa of the intersection point of
relaxation curve and this dashed line.
In Fig. 6, the relaxation time for various local potential $U_0$ is presented
at $T > T_c$.
Here we adopt two legible parameters, $\delta$ and $\xi$, to describe the 
potential wells and barriers for $U_0$ (see Fig. 3), which are defined by
\begin{eqnarray}
\delta & \equiv & [ U_0 (x_b) - U_0 (x_c) ] / \omega_0 , \\
\xi & \equiv & U_0 (x_c) / \omega_0 .
\end{eqnarray}
Provided $\delta$ and $\xi$, $c_4$ and $c_6$ can be derived in terms
of Eqs. (35)-(38).
The values of $c_4$ and $c_6$ for the calculation of Fig. 6 are listed in
Table I, where we set $\xi$=3.061 and change $\delta$ from 4.239 to 4.639.
The leftmost point on each curve denotes the $t_r$ at just above
$T_c$, which is a temperature determined from the singular point of
$C_f^V$ according to Eq. (28).

\begin{figure}
\begin{center}
\includegraphics{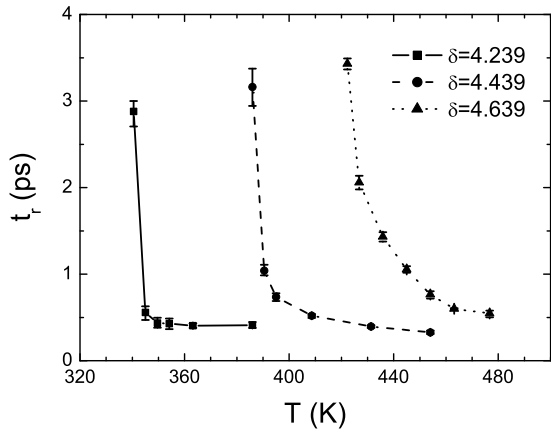}
\end{center}
\caption{
Temperature dependence of relaxation time $t_r$ for various $\delta$ when
$T > T_c$, where $\xi$ is fixed at 4.439.
Error bars show the 95\% confidence interval.
}
\end{figure}

As revealed by the NMR experiment,\cite{za03} the paraelectric-ferroelectric
phase transition of BaTiO$_3$ has both displacive and order-disorder
components in its mechanism.
Short range dipole fluctuation arises in the paraelectric phase near $T_c$
as a precursor of the order-disorder transition, and condenses into long
range ferroelectric ordering below $T_c$.
Thus, in the present study, the relaxation of photo-created cluster is
also subject to the dynamics of this dipole fluctuation and
yields a temperature dependence.
As illustrated by the three curves in Fig. 6, if a ferroelectric cluster
is created at a temperature close to $T_c$, relaxation of this cluster
is slow because of a rather strong dipole fluctuation, which holds the
cluster in the metastable ferroelectric state from going back
to the paraelectric one.
Away from $T_c$, $t_r$ decreases considerably for the dipole fluctuation
is highly suppressed.
This behavior is nothing but the critical slowing down of photorelaxation.

\begin{table}
\caption{
Parameters adopted for calculation of Fig. 6.
}
\begin{ruledtabular}
\begin{tabular}{ccccc}
$c_4$ & $c_6$ & $\delta$ & $\xi$ & $T_c$ (K)\\ 
\hline
2.0696$\times$10$^{-2}$ & 3.4521$\times$10$^{-4}$ & 4.239 & 3.061 & 340\\
2.0132$\times$10$^{-2}$ & 3.2593$\times$10$^{-4}$ & 4.439 & 3.061 & 386\\
1.9596$\times$10$^{-2}$ & 3.0814$\times$10$^{-4}$ & 4.639 & 3.061 & 422\\
\end{tabular}
\end{ruledtabular}
\end{table}

In Fig. 6, it can also be seen that with the increase of $\delta$, $T_c$
moves to the high temperature side so as to overcome a higher potential
barrier between the ferroelectric and paraelectric phases.
Furthermore, the evolution of $t_r$ becomes gentle as well, implying a
gradual weakening of dipole fluctuation at high temperature region.

\begin{figure}
\begin{center}
\includegraphics{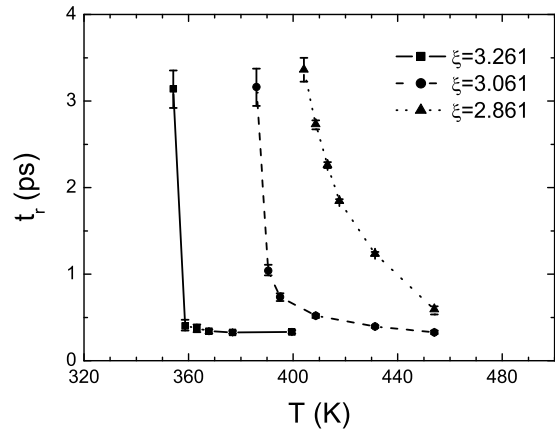}
\end{center}
\caption{
Temperature dependence of relaxation time $t_r$ for various $\xi$ when
$T > T_c$, where $\delta$ is fixed at 3.061.
Error bars show the 95\% confidence interval.
}
\end{figure}

\begin{table}
\caption{
Parameters adopted for calculation of Fig. 7.
}
\begin{ruledtabular}
\begin{tabular}{ccccc}
$c_4$ & $c_6$ & $\delta$ & $\xi$ & $T_c$ (K)\\ 
\hline
1.9626$\times$10$^{-2}$ & 3.1070$\times$10$^{-4}$ & 4.439 & 3.261 & 354\\
2.0132$\times$10$^{-2}$ & 3.2593$\times$10$^{-4}$ & 4.439 & 3.061 & 386\\
2.0663$\times$10$^{-2}$ & 3.4223$\times$10$^{-4}$ & 4.439 & 2.861 & 404\\
\end{tabular}
\end{ruledtabular}
\end{table}

In Fig. 7, we show the temperature dependence of $t_r$ for different $\xi$
when $T > T_c$, where $\delta$ is fixed at 4.439.
The values of parameters for this calculation are given in Table II.
When $\xi$ changes from 3.261 to 2.861, as shown in Fig. 7, $T_c$
gradually increases.
This is because with the decrease of $\xi$, the ferroelectric state at
$x_c$ (refer to Fig. 3) becomes more stable and can survive even larger
thermal fluctuation.
In a manner analogous to Fig. 6, the evolution of $t_r$ also displays
a sharp decline at low temperature, and becomes more and more smooth
as temperature increases.

\begin{figure}
\begin{center}
\includegraphics{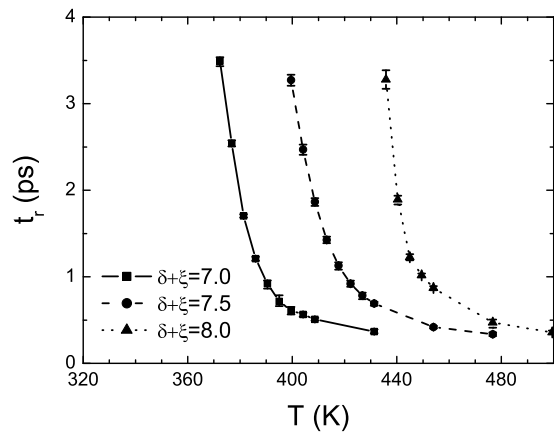}
\end{center}
\caption{
Temperature dependence of relaxation time $t_r$ for various barrier height
$\delta + \xi$ when $T > T_c$, where $\delta / \xi$=1.5 is assumed.
Error bars show the 95\% confidence interval.
}
\end{figure}

In Fig. 8, we plot the temperature dependence of $t_r$ for different barrier
heights, {\it i.e.}, $\delta + \xi$ varies from 7.0 to 8.0, while the
ratio $\delta/\xi$ is fixed at 1.5.
Parameters for this calculation are provided in Table III.
As already discussed with Figs. 6 and 7, larger $\delta$ tends to increases
$T_c$, but higher $\xi$ applies an opposite effect on $T_c$.
Combining these two effects, in Fig. 8, one finds that $T_c$ increases
if both $\delta$ and $\xi$ are enhanced, indicating that in this case,
the change of $\delta$ plays a more significant role.
Meanwhile, in contrast to Figs. 6 and 7, all the three curves in Fig. 8
present smooth crossovers on decreasing temperature towards $T_c$, signifying
that dipole fluctuation can be promoted by lowering $\xi$ even the
temperature is decreased.

\begin{table}
\caption{
Parameters adopted for calculation of Fig. 8.
}
\begin{ruledtabular}
\begin{tabular}{cccccc}
$c_4$ & $c_6$ & $\delta$ & $\xi$ & $\delta + \xi$ & $T_c$ (K)\\
\hline
2.1557$\times$10$^{-2}$ & 3.7309$\times$10$^{-4}$ & 4.200 & 2.800 & 7.000 & 372\\
2.0122$\times$10$^{-2}$ & 3.2505$\times$10$^{-4}$ & 4.500 & 3.000 & 7.500 & 400\\
1.8860$\times$10$^{-2}$ & 2.8557$\times$10$^{-4}$ & 4.800 & 3.200 & 8.000 & 436\\
\end{tabular}
\end{ruledtabular}
\end{table}

In Namikawa's experiment, the wavelength of soft x-ray laser is 160 {\AA},
hence the photo-created cluster is of nano scale.
However, it should be noted that relaxation of nano-sized cluster is beyond
our present quantum Monte Carlo simulation because of the size limitation
of our model.
This is the primary reason why the experimentally measured relaxation
time can reach about 30 picoseconds, being several times longer than
our calculated results.
In spite of the difference, our calculation has well clarified the critical
dynamics of BaTiO$_3$ and the origin of speckle variation.

\section{Summary}

We carry out a theoretical investigation to clarify the dynamic property
of photo-created ferroelectric cluster observed in the paraelectric
BaTiO$_3$ as a real time correlation of speckle pattern between two soft
x-ray laser pulses.
The density matrix is calculated by a perturbative expansion up to the
fourth order terms, so as to characterize the time dependence of
scattering probability.
The cluster-associated phonon softening as well as central peak effects
are well reproduced in the phonon spectral function via a quantum Monte
Carlo simulation.
We show that the time dependence of speckle pattern is determined by the
relaxation dynamics of photo-created ferroelectric cluster, which is
manifested as a central peak in the phonon spectral function.
The photorelaxation of ferroelectric cluster is featured by a critical
slowing down on decreasing the temperature.
Near the $T_c$, cluster excitation is stablized by the strong dipole
fluctuation, correspondingly the relaxation becomes slow.
While, at higher temperature, dipole fluctuation is suppressed, ending
up with a quicker relaxation of cluster.
Our simulation also illustrates that the critical slowing down and dipole
fluctuation are subject to the chemical environment of crystal.

\section{Acknowledgments}
This work is supported by the Next Generation Supercomputer Project,
Nanoscience Program, MEXT, Japan.


\begin{thebibliography}{99}

\bibitem{ma77} M. May,
  J. Phys. E \textbf{10} 849 (1977).

\bibitem{go07} J. W. Goodman,
  {\it Speckle Phenomena in Optics: Theory and Applications}
  (Roberts and Company, Greenwood Village, 2007).

\bibitem{ta02} R. Z. Tai, K. Namikawa, M. Kishimoto, M. Tanaka, K. Sukegawa,
  N. Hasegawa, T. Kawachi, M. Kado, P. Lu, K. Nagashima, H. Daido,
  H. Maruyama, A. Sawada, M. Ando, and Y. Kato,
  Phys. Rev. Lett. \textbf{89}, 257602 (2002).

\bibitem{ta04} R. Z. Tai, K. Namikawa, A. Sawada, M. Kishimoto, M. Tanaka,
  P. Lu, K. Nagashima, H. Maruyama, and M. Ando,
  Phys. Rev. Lett. \textbf{93}, 087601 (2004).

\bibitem{ha95} S.-I. Hamazaki, F. Shimizu, S. Kojima, and M. Takashige,
  J. Phys. Soc. Jpn. \textbf{64} 3660 (1995).

\bibitem{pa98} G. K. H. Pang and K. Z. Baba-Kishi,
  J. Phys. D \textbf{31} 2846 (1998).

\bibitem{ya69} Y. Yamada, G. Shirane, and A. Linz,
  Phys. Rev. \textbf{177} 848 (1969).

\bibitem{mu96} W. L. Mulvihill, K. Uchino, Z. Li and W. Cao,
  Phil. Mag. B \textbf{74} 25 (1996).

\bibitem{po98} D. L. Polla and L. F. Francis,
  Annu. Rev. Mater. Sci. \textbf{28} 563 (1998).

\bibitem{ha71} J. Harada, J. D. Axe, and G. Shirane,
  Phys. Rev. B \textbf{4} 155 (1971).

\bibitem{mi76} R. Migoni, D. Bauer, and H. Bilz,
  Phys. Rev. Lett. \textbf{37} 1155 (1976).

\bibitem{za03} B. Zalar, V. V. Laguta, and R. Blinc
  Phys. Rev. Lett. \textbf{90} 037601 (2003).

\bibitem{vo07} G. V\"{o}lkel and K. A. M\"{u}ller,
  Phys. Rev. B \textbf{76} 094105 (2007).

\bibitem{na08} K. Namikawa, unpublished.

\bibitem{ya08} R. Yan, Z. Guo, R. Tai, H. Xu, X. Zhao, D. Lin, X. Li, and H. Luo,
  Appl. Phys. Lett. \textbf{93}, 192908 (2008).

\bibitem{kr75} J. A. Krumhansl and J. R. Schrieffer,
  Phys. Rev. B \textbf{11}, 3535 (1975).

\bibitem{au75} S. Aubry,
  J. Chem. Phys. \textbf{62} 3217 (1975).

\bibitem{sc76} T. Schneider and E. Stoll,
  Phys. Rev. B \textbf{17}, 1302 (1978).

\bibitem{sa02} V. V. Savkin, A. N. Rubtsov, and T. Janssen,
  Phys. Rev. B \textbf{65} 214103 (2002).

\bibitem{mo90} J. R. Morris and R. J. Gooding,
  Phys. Rev. Lett. \textbf{65}, 1769 (1990).

\bibitem{kh08}  A. Khare, A. Saxena,
  J. Math. Phys. \textbf{49}, 063301 (2008).

\bibitem{na94} K. Nasu,
  J. Phys. Soc. Jpn. \textbf{63}, 2416 (1994).

\bibitem{do74} S. Doniach and E. H. Sondheimer,
  {\it Green's Function for Solid State Physicists},
  (Benjamin, London, 1974), Appendix 2.

\bibitem{ji04} K. Ji, H. Zheng, and K. Nasu,
  Phys. Rev. B \textbf{70}, 085110 (2004).

\bibitem{bo96} J. Bon\v{c}a and J. E. Gubernatis,
  Phys. Rev. B \textbf{53}, 6504 (1996).

\bibitem{sk84} Skilling and R. K. Bryan,
  Mon. Not. R. Astron. Soc. \textbf{211}, 111 (1984).

\bibitem{ya03} M. Yamazaki, N. Tomita, and K. Nasu,
  J. Phys. Soc. Jpn. \textbf{72}, 611 (2003).

\bibitem{gu91} J. E. Gubernatis, M. Jarrell, R. N. Silver, and D. S. Sivia
  Phys. Rev. B \textbf{44}, 6011 (1991).

\bibitem{zh94} W. Zhong, R. D. King-Smith, and D. Vanderbilt,
  Phys. Rev. Lett. \textbf{72}, 3618 (1994).


\end{thebibliography}
\end{document}